\documentstyle[12pt]{article}

\newcommand{\be}{\begin{equation}}
\newcommand{\ee}{\end{equation}}
\newcommand{\bear}{\begin{eqnarray}}
\newcommand{\ear}{\end{eqnarray}}
\newcommand{\lsi}{\raise0.3ex\hbox{$<$\kern-0.75em\raise-1.1ex\hbox{$\sim$}}}
\newcommand{\gsi}{\raise0.3ex\hbox{$>$\kern-0.75em\raise-1.1ex\hbox{$\sim$}}}
\newcommand{\lsim}{\mathop{\lsi}}

\newcommand{\newsubsection}[1]{
\vspace{1cm}
\pagebreak[3] \addtocounter{subsection}{1}
\nopagebreak
\vspace{2mm}
\nopagebreak}

\begin{document}

\begin{center}\vspace*{1.0cm}
{\Large\bf Instantons and spontaneous } \\ {\Large\bf
color symmetry breaking} \\
\vspace*{1.0cm} {\large Christof Wetterich} \\
\vspace*{0.5cm}{\normalsize {Institut f\"ur Theoretische Physik, \\
Universit\"at Heidelberg, \\ Philosophenweg 16, \\ D--69120 Heidelberg,
Germany.}} \\

\end{center}

\begin{abstract}
The instanton interaction in QCD generates an effective
potential for scalar quark-antiquark condensates in the color
singlet and octet channels. For three light quark flavors the
cubic term in this potential induces an octet condensate and
``spontaneous breaking'' of color in the vacuum. Realistic
masses of the $\rho$- and $\eta'$-mesons are compatible with
renormalization-group-improved instanton perturbation theory.
\end{abstract}
It has been argued recently \cite{GM,Mean} that the physics of
confinement in long distance QCD admits an equivalent description in the
Higgs picture where color is ``spontaneously broken'' by an octet
quark-antiquark condensate
\bear\label{1}
&&<\bar\psi_{Ljb}\psi_{Rai}>-\frac{1}{3}<\bar\psi_{Lkb}\psi_{Rak}>
\delta_{ij}=\nonumber\\
&&\frac{1}{\sqrt6}\bar\xi(\delta_{ia}\delta_{jb}-\frac{1}{3}\delta_{ij}
\delta_{ab})
\ear
The structure in the color indices $i,j,k=1...3$ and flavor indices
$a,b=1...3$ is such that a physical $SU(3)$ symmetry remains
unbroken. The physical vector meson states (gluons) transform as an
octet with an equal mass $\sim \bar\xi$. They have integer electric
charge and can be associated with the $\rho-,K^*-$
and $\omega$-mesons. Also the fermions (quarks) transform
as a massive octet (plus a heavy singlet) with the appropriate charges
to describe the baryon octet $(p,n,\Lambda,\Sigma,\Xi)$. (We consider
here three  flavors of light quarks and neglect the $SU(3)$-splitting
in vacuum expectation values due to the mass of the
strange quark.) The Higgs mechanism generates a mass for the gluons
and therefore provides for an effective infrared cutoff in QCD. This also
gives a simple explanation for the confinement of color charges: the
gauge fields between such charges are squeezed into flux tubes by
effect of the mass, in analogy to the Meissner effect in
superconductors. Furthermore, a simple effective action for scalars
representing quark-antiquark-bound states $\sim\bar\psi\psi$
leads to a very successful phenomenological picture, including realistic
pion-nucleon couplings, vector-dominance for the electromagnetic
interactions of pions, realistic decay rates of the $\rho$-mesons
into pions and charged leptons and an explanation of the $\Delta I=
1/2$ rule for weak hadronic kaon decays \cite{GM}.

We propose in this letter a simple dynamical mechanism how spontaneous
color breaking is generated in QCD. It is based on the effective
't Hooft interaction for instanton effects \cite{Hooft}, \cite{Shif}, in accordance
with speculations that instantons are crucial for an understanding
of low energy QCD \cite{Instanton,5a}. In short, the instanton-induced axial
anomaly induces a ``cubic term'' in the effective potential
for scalar $\bar\psi\psi$-states which drives the minimum both for
color octet and singlet scalars away from $<\bar\psi\psi>=0$.

Besides a dynamical explanation of the color octet condensate our
approach also solves two important problems in instanton physics.

(1) First the old question about the effective infrared cutoff
for very large size QCD instantons is answered by the colored
octet condensate.
The induced  gluon mass acts as a cutoff, very similar
to the $W$-boson mass for electroweak instantons. Furthermore, the
physics of confinement is now  integrated in the instanton
physics.

(2) Second we solve the problem of the ``unboundedness of the naive
instanton interaction'', which we explain briefly in
the following. For three massless flavors the contribution
of instantons with size $\rho$ to the effective $U(1)_A$-violating
fermion interaction reads\footnote{We take the opportunity to correct
the instanton vertex of \cite{Mean}. It was based on the Fierz transform
of an uncorrect vertex quoted in \cite{5a}.}
\cite{Hooft}, \cite{Shif}
\bear\label{2}
d{\cal L}&=&-d\zeta(\rho){\cal A}\\
{\cal A}&=&\det\tilde\varphi^{(1)}+\det
\tilde\varphi^{(2)}
-\frac{3}{4}(E(\tilde\varphi^{(1)},\tilde\chi^{(1)})+
E(\tilde\varphi^{(2)},
\tilde\chi^{(2)}))\nonumber\ear
with quark-antiquark bilinears
\bear\label{3}
\tilde\varphi^{(1)}_{ab}&=&\bar\psi_{L\ ib}\ \psi_{R\ ai}\quad,\quad
\tilde\varphi^{(2)}_{ab}=-\bar\psi_{R\ ib}\ \psi_{L\ ai}
\nonumber\\
\tilde\chi^{(1)}_{ij,ab}&=&\bar\psi_{L\ jb}\ \psi_{R\ ai}
-\frac{1}{3}\bar\psi_{L\ kb}\ \psi_{R\ ak}\ \delta_{ij}\nonumber\\
\tilde\chi^{(2)}_{ij,ab}&=&-\bar\psi_{R\ jb}\ \psi_{L\ ai}
+\frac{1}{3}\bar\psi_{R\ kb}\ \psi_{L\ ak}\ \delta_{ij}\ear
and
\be\label{4}
E(\tilde\varphi,\tilde\chi)=\frac{1}{6}\epsilon_{a_1a_2a_3}
\epsilon_{b_1b_2b_3}\tilde\varphi_{a_1b_1}\tilde\chi_{ij,a_2b_2}
\tilde\chi_{ji,a_3b_3}\ee
Partial bosonization\footnote{Partial bosonization requires that the
effective scalar potential is bounded from below (for details see
\cite{Mean}). We will see that this is indeed the case.}
replaces  the quark-antiquark bilinears  (\ref{3})
by appropriate
singlet and octet bosonic fields $\tilde\varphi^{(1)}_{ab}\to\sigma_{ab},
\tilde\varphi^{(2)}_{ab}\to\sigma^\dagger_{ab},\tilde\chi^{(1)}_{ij,ab}
\to\xi_{ijab},\tilde\chi^{(2)}_{ijab}\to\xi_{ji,ba}^*$. Correspondingly,
the interaction (\ref{2}) transmutes into an effective potential for the
scalar fields. An evaluation along  the directions (\ref{1}) and
$<\bar\psi_{Lib}\psi_{Ria}>=\bar\sigma\delta_{ab}$
results for constant $\zeta$ in a cubic effective
potential
\be\label{5}
U_{an}(\bar\sigma,\bar\xi)=-\zeta(2\bar\sigma^3+
\frac{1}{3}\bar\sigma\bar\xi^2)\ee
It is obvious that for $\bar\sigma>0$
the effective potential can always be arbitrarily
lowered by an increase of the color octet condensate $|\bar\xi|$.
We will see that eq. (\ref{5}) with constant  $\zeta$ is a
valid approximation for not too large $|\bar\xi|$ and conclude
that the
instanton interaction induces spontaneous
color symmetry breaking. The problem
is that the effective instanton potential (\ref{5}) is unbounded
for large $\bar\sigma^2,\bar\xi^2$. Even though there are,
in principle, stabilizing higher-order interactions $\sim\bar
\sigma^4,\bar\xi^4$ induced by anomaly-free loop graphs with
eight quark/antiquark legs, it makes no sense that the instanton
contribution to the effective action increases without bounds for
large values of the chiral condensates.

The approximation of $d\zeta(\rho)$
being independent of $\tilde\varphi$ and $\tilde \chi$ or,
equivalently, $\bar\sigma$ and $\bar\xi$, holds only for small values
of the chiral condensate (e.g. $|\bar\sigma\rho^3|\simeq
3/(2\pi^2)$ \cite{Shif}). The behavior for large values of
the condensates is dominated by two effects. (a) For large $\bar\xi^2$
the gluons become massive and cut off the instanton contribution.
(b) For large $\bar\sigma^2$ or $\bar\xi^2$ the quarks become heavy
and the  dependence
of the instanton interaction on $\bar\sigma$ and $\bar\xi$
disappears. We concentrate here on the
first aspect. Instead of a careful
study of the dependence of $d\zeta(\rho)$ on the gluon mass and therefore
on $\bar\xi$ we take a simplified approach which reflects the
qualitative behavior correctly: we neglect the influence of $\bar\xi$
for small $\rho$ and omit the suppressed contribution for large $\rho$.
As a result, the  coefficient $\zeta$ depends on the
value of the color octet condensate $\bar\xi$ by the appearance  of an
effective cut-off $\rho_{\rm max}(\bar\xi)$ in the integral over
instanton sizes.

The effective instanton vertex \cite{Hooft}
is therefore multiplied by
\be\label{6}
\zeta(\bar\xi)=\frac{32}{15}\pi^6\kappa
(f'(0))^3C_3\int^{\rho_{\rm max}(\bar\xi)}_0
\frac{d\rho}{\rho} f(\rho)\ee
\be\label{7}
f(\rho)=\rho^5\left(\frac{\alpha(1/\rho)}
{\alpha(\bar\mu)}\right)^{-\frac{4}{3}}
\left(\frac{2\pi}{\alpha(1/\rho)}\right)^6\exp
\left(-\frac{2\pi}{\alpha(1/\rho)}\right)
\ee
Here $\alpha(\mu)=4\pi g^2(\mu)$ corresponds to the running coupling in
the $\overline{\rm MS}$-scheme in three-loop order \cite{PDG}
with  $\Lambda=\Lambda^{(3)}_{\overline{MS}}=330$ MeV. The
prefactor $C_3$ reads for the $\overline{MS}$-scheme $C_3=1.51\cdot10^{-3}$
and $f'(0)=1.34$. We note that the height
of the maximum of $f(\rho)$ depends  on the precise
definition of $\alpha$ and its $\beta$-function. We have therefore
introduced in (\ref{6}) a constant $\kappa$ of order one which
parametrizes this uncertainty. The perturbative value is $\kappa=1$.
Furthermore, $\kappa$ accounts for the ambiguity in the Fierz
transformation of the instanton vertex.
We adopt the normalization scale for the fermion operators $\bar\mu=2$
GeV. Due to the strong increase of $\alpha(1/\rho)$ the function
 $f(\rho)$ vanishes for $\rho\to 1/\Lambda$ and we may take
$f(\rho)\equiv 0$ for $\rho>1/\Lambda$. The maximum of $f(\rho)$
at $\rho^{-1}=613$ MeV is not very far from the ``perturbative range''.
For the range
$\rho^{-1}\geq 800$ MeV the approximation  (\ref{6}) may be considered
as a reliable guide, whereas for $\rho^{-1}\leq$ 500 MeV it
is expected \cite{Shif} to break down. It seems reasonable
to believe the
qualitative feature of eqs. (\ref{6}), (\ref{7}),
 namely that $f(\rho)$ suppresses the
contribution of very large instantons such that $\zeta$ remains finite
for $\rho_{\rm max}\to\infty$. This is, however, not crucial for our
argument.

The effective cutoff $\rho_{\rm max}(\bar\xi)$ is proportional to the
inverse of the $\bar\xi$-dependent effective gauge boson mass
$\mu_\rho(\bar\xi)$. By
the Higgs mechanism the
effective gauge boson mass is, in turn,
proportional to the octet condensate
 $\bar\xi$ and the effective gauge coupling
$g(\mu_\rho)$
\be\label{8b}
\mu^2_\rho(\bar\xi)=g^2(\mu_\rho) Z\bar\xi^2\ee
We find it convenient to use $\mu_\rho$ instead of $\bar\xi$
as the
independent variable. Inverting the functional dependence
 $\bar\xi(\mu_\rho)=Z^{-1/2}\mu_\rho/g(\mu_\rho)$ one obtains a lower bound for
$\mu_\rho$
\be\label{9}
\mu_\rho(\bar\xi=0)=\Lambda\quad,\quad g^2(\bar\xi\to 0)\sim
\Lambda^2/\bar\xi^2\ee
The unknown details of the way how the gluon mass acts as an
infrared cutoff are absorbed into a proportionality factor
$c_\rho$ of order one
\be\label{9a}
\rho_{max}(\bar\xi)=c_\rho/\mu_\rho\ee
Inserting this cutoff in eq. (\ref{6}), one finds that for
$\bar\xi\to 0$ the coefficient $\zeta(\bar\xi)$ becomes almost
independent of $\bar\xi$ whereas for large $\bar\xi$ it decreases
rapidly  $\sim\bar\xi^{-14}$. This qualitative behavior is sufficient
for instanton induced color symmetry breaking, independent of the
quantitative details. Indeed, the potential vanishes for $\bar\xi=0$ and  
$|\bar\xi|\to\infty$ and takes negative values in a range of
finite nonzero $\bar\xi$. For small values of $|\bar\xi|$ and
arbitrary nonzero positive $\bar\sigma$ the term $\sim  
-\zeta\bar\sigma\bar\xi^2$ acts like a negative mass term
for $\bar\xi$ which destabilizes the line $\bar\xi=0$.

On the other hand, for fixed $\bar\xi$ and $\bar\sigma^2\to\infty$ all
effective fermion masses diverge and the instanton contribution
depends on the fermion bilinears only through the effective
gluon mass or $\rho_{max}(\bar\xi)$. The
potential becomes positive
\bear\label{5a}
&&\lim_{\bar\sigma^2\to\infty} U_{an}(\bar\sigma,\bar\xi)
=\nonumber\\
&&C_3\int^{\rho_{max}}_0d\rho\rho^{-5}\left(\frac{2\pi}{\alpha(1/\rho)}
\right)^6\exp\left(-\frac{2\pi}{\alpha(1/\rho)}\right)\ear
and this guarantees the boundedness in the $\bar\sigma$-direction.

We are now ready to discuss the instanton potential quantitatively
by replacing in eq. (\ref{5}) $\zeta\to\zeta(\bar\xi)$.
In our conventions $\zeta$ is positive and the minimum in the
 $\bar\sigma$-direction occurs for $\bar\sigma_0\geq 0$. The determination
of the ratio of expectation values $r_0=\bar\sigma_0/\bar\xi_0$
depends on details of the stabilization of the potential in the
$\bar\sigma$-direction which we do not investigate here. We keep
$r_0$ as a parameter and investigate the potential on the
line $\bar\sigma=r_0\bar\xi$
\be\label{14}
\bar U_{an}(\mu_\rho)=-\frac{r_0}{3}(1+6r_0^2)Z^{-\frac{3}{2}}
\frac{\mu^3_\rho}{g^3(\mu_\rho)}\zeta(\mu_\rho)\ee
The location of the minimum
is independent of $r_0,Z$ and the prefactor $\kappa$
multiplying the integral
(\ref{6}). The vacuum expectation value $\bar\mu_\rho$ depends, however,
on the unknown constant $c_\rho$, as shown in the table (with mass unit GeV).
The only scale is set by the perturbative running of $\alpha$
and therefore $\bar\mu_\rho\sim\Lambda$.

\begin{center}
\begin{tabular}{|c|c|c|c|c|}
 $c_\rho$&$\bar\mu_\rho$&$\rho^{-1}_{\rm max}(\bar\mu_\rho)$&
$|<\bar\psi\psi>|^{1/3}$&$\zeta_0/\kappa$\\
\hline
1.0&0.65&0.65&0.28&223\\
1.2&0.74&0.62&0.27&304\\
1.4&0.84&0.60&0.27&343\\
\end{tabular}\\
\end{center}

\medskip\noindent
{\bf Table:}
Vector-meson mass $\bar\mu_\rho$ and singlet chiral
condensate $<\bar\psi\psi>$.

This establishes our main result, namely that the instanton interaction
leads to spontaneous color  symmetry breaking, with expectation value
$\bar\xi_0=
\bar\xi(\bar\mu_\rho)\not= 0!$ The value $\bar\mu_\rho$ should be
identified with the mass of the $\rho$ and $K^*$ mesons  in the limit
of a vanishing strange quark mass $m_s$. In view of the uncertainties
the result 600 MeV$\lsim\bar\mu_\rho\lsim 900$ MeV is very satisfactory.
It should motivate a more detailed study how the gluon mass term cuts
off the instanton integral, which
amounts to a computation of $c_\rho$. (Since fluctuations with
$\rho^{-2}<\mu^2_\rho$ are only supressed instead of being completely
eliminated we expect $c_\rho\geq 1$). We observe that $\rho_{\rm max}$
is typically near the maximum of $f(\rho)$ (eq.(\ref{7})).

We next turn to the mass of the $\eta'$-meson which is directly
related to the value of the anomaly potential at the minimum
\cite{GM}
\be\label{13}
M^2_{\eta'}-\tilde m^2_g=-\frac{18}{f^2_\theta}U_{an}(\bar\sigma_0,
\bar\xi_0)=\frac{36\zeta_0}{f^2_\theta}(\bar\sigma_0^3
+\frac{1}{6}\bar\sigma_0\bar\xi_0^2)\ee
For $f_\theta$ we use \cite{GM} $f^2_\theta=f^2
(16x+7)/(7(1+x))$ which yields for the average meson decay
constant $f=106$ MeV and large $x$ (i.e. $x=6$)
$f_\theta=$ 154 MeV. (The corresponding two-photon
decay width of the $\eta'$, i. e. $\Gamma(\eta'\to2\gamma)=
\alpha^2_{em}M^3_{\eta'}/(24\pi^3 f^2_\theta)=2.7$ keV, agrees
reasonably well with observation.) We also account here for the
effect of nonvanishing quark masses \cite{PDG} by
$\tilde m_g=410 {\rm MeV} \sqrt{Z_m/Z_p}=$ 340 MeV. In the table we
show the singlet quark-antiquark condensate $\bar\sigma_0
=-\frac{1}{2}<\bar\psi\psi>(\bar\mu= 2 {\rm GeV})$ which corresponds
to $M_{\eta'}=$ 960 MeV for $r_0=0.5$ and $\kappa=1$. (Typical values
of $|<\bar\psi\psi>|^{1/3}$ for $r_0=1/3$ are 10 MeV smaller.)
The results agree well with common estimates. We conclude that
our estimate of $\zeta_0=\zeta(\bar\mu_\rho)$ leads to
a realistic value of $M_{\eta'}$!

We finally point out that the contributions of the nonvanishing
strange quark mass to the effective potential are of similar magnitude
as the anomaly-induced potential. They tend to increase $\bar\sigma_0$.
It is interesting to note that the anomaly in two-flavor QCD has
also the tendency to induce an octet condensate due to a negative
quadratic term for $\bar\xi$.

In presence of a (not too large) strange quark mass  and the four fermion
interactions generated by QCD-box diagrams the effective potential acquires
additional contributions\cite{Mean}
\bear\label{V1}
&&\Delta U(\bar\sigma,\mu_\rho)=-2m_s\bar\sigma-
\zeta^{(s)}(\mu_\rho)m_s(2\bar\sigma^2+\frac{1}{6}
\bar\xi^2(\mu_\rho))\nonumber\\
&&+\frac{g^4(\mu_\rho)}{16\pi^2\mu^2_\rho}\left(\frac{g^2(\mu_\rho)}
{g^2(\bar\mu)}\right)^{-\frac{8}{9}}(
3L_\sigma\bar\sigma^2+\frac{4}{3}L_\chi\bar\xi^2(\mu_\rho))\ear
The gluon mass acts as cutoff in the box diagrams\footnote{For a detailed
renormalization group investigation of the one-particle irreducible
four-quark interactions $\sim\bar\sigma^2$ see \cite{Enrico}.}
and we neglect the effect of the fermion mass, resulting in
$L_\sigma=\frac{552}{117} L_\chi=\frac{23}{9}$.
We note $d\zeta^{(s)}=(5/6)(\pi\rho)^{-2}
(\alpha(1/\rho)/\alpha(\bar\mu))^{8/9}d\zeta$ and an additional
contribution to the $\eta'$-mass $\Delta M^2_{\eta'}=16m_s\zeta^{(s)}(\bar
\mu_\rho)\cdot(\bar\sigma_0^2+\bar\xi^2_0/12)/f^2_\theta$.

In conclusion, the instanton-induced anomalous six-quark interactions
induce the spontaneous breaking of color. The destabilization of the
``color-symmetric state'' $\bar\xi=0$ due to the instanton
interaction seems to be quite robust. We could not find competing stabilizing
terms from other effective interactions in QCD. The pure instanton
interaction with fluctuations evaluated perturbatively and
quark mass effects neglected gives already a very satisfactory picture
with a realistic range for the vector-meson masses and the mass of
the $\eta'$-meson.

\noindent {\bf Acknowledgement:} The author would like to thank J.
Berges for stimulating discussions.

\end{document}